\documentclass[3p,times,twocolumn]{elsarticle}

\usepackage{ecrc}

\volume{00}

\firstpage{1}

\journalname{Nuclear Physics B Proceedings Supplement}

\runauth{Shun Zhou}

\jid{nuphbp}

\jnltitlelogo{Nuclear Physics B Proceedings Supplement}

\usepackage{amssymb}

\usepackage[figuresright]{rotating}

\begin{document}

\begin{frontmatter}



\dochead{}

\title{Overview on Neutrino Theory and Phenomenology}


\author[IHEP,CHEP]{Shun Zhou\fnref{email}}
\fntext[email]{Email address: zhoush@ihep.ac.cn}

\address[IHEP]{Institute of High Energy Physics, Chinese Academy of Sciences, Beijing 100049, China}
\address[CHEP]{Center for High Energy Physics, Peking University, Beijing 100871, China}
\begin{abstract}
In this talk, I give an overview on recent theoretical and phenomenological studies of massive neutrinos. First of all, the present status of neutrino mixing parameters is summarized. The phenomenology of neutrino oscillations is then discussed, and current understanding of lepton flavor mixing is presented. Finally, I consider the seesaw models of neutrino masses and briefly mention the direct searches at the CERN Large Hadron Collider (LHC).
\end{abstract}


\end{frontmatter}


\section{Introduction}
\label{intro}

This year celebrates the 60th anniversary of the discovery of neutrinos, which were introduced by Pauli in 1930 to save energy conservation in $\beta$ decays~\cite{Pauli:1930pc}. Since electron antineutrinos $\overline{\nu}^{}_e$ from nuclear reactors were first detected by Cowan and Reines in 1956~\cite{Cowan:1992xc}, many great achievements have been made in neutrino physics~\cite{Xing:2011zza}:
\begin{itemize}
\item {\it Transitions of neutrino flavors} -- In 1957, Pontecorvo~\cite{Pontecorvo:1957qd} considered the transition $\overline{\nu}^{}_e \to \nu^{}_e$ in an analogy with the $K^0$-$\bar{K}^0$ mixing~\cite{GellMann:1955jx}. Furthermore, Maki, Nakagawa and Sakata conjectured in 1962 a possible flavor transition $\nu^{}_e \to \nu^{}_\mu$ and realized it in a theoretical model of elementary particles~\cite{Maki:1962mu}.

\item {\it Three flavors of neutrinos} -- In 1962, the second flavor of neutrino $\nu^{}_\mu$ was discovered~\cite{Danby:1962nd}. Although the third flavor $\nu^{}_\tau$ was expected to exist after the discovery of the charged leptons $\tau^\pm$ in 1975~\cite{Perl:1975bf}, it was finally observed in 2000 by the DONUT Collaboration~\cite{Kodama:2000mp}. This late discovery indicates that the flavor puzzle exists not only for quarks but also for lepotons.

\item {\it Neutrino oscillations} -- Following a brilliant idea from Chen~\cite{Chen:1985na}, the SNO Collaboration perfectly solved the problem of missing solar neutrinos $\nu^{}_e$ in 2002 by measuring both $\nu^{}_e$ and neutrinos of non-electron flavors~\cite{Ahmad:2002jz}. On the other hand, the Super-Kamiokande Collaboration provided strong evidence for the disappearance of atmospheric neutrinos $\nu^{}_\mu$ and $\overline{\nu}^{}_\mu$ in 1998~\cite{Fukuda:1998mi}. The observation of reactor neutrino disappearance in 2012 by the Daya Bay Collaboration~\cite{An:2012eh} leads to a complete picture of three-flavor neutrino oscillations, involving three mixing angles $\{\theta^{}_{12}, \theta^{}_{23}, \theta^{}_{13}\}$ and two neutrino mass-squared differences $\{\Delta m^2_{21}, \Delta m^2_{31}\}$ with $\Delta m^2_{ji} \equiv m^2_j - m^2_i$ for $ji = 21, 31$.
\end{itemize}
The discovery of neutrino oscillations was awarded the Nobel Prize in Physics in 2015~\cite{Kajita:2016cak,McDonald:2016ixn}, and the most important experimental collaborations for neutrino oscillations and their leaders shared the Breakthrough Prize in Fundamental Physics in 2016. This recognition marks the end of a golden time of neutrino physics since 1998, and starts a new era of precision measurements of neutrino oscillation parameters and a more profound understanding of neutrino mass generation and lepton flavor mixing~\cite{pdg,Zhou:2015kqs}.

\section{Current Status}
\label{status}

The phenomena of neutrino oscillations and lepton flavor mixing can be described by the Pontecorvo-Maki-Nakagawa-Sakata (PMNS) matrix~\cite{Pontecorvo:1957qd,Maki:1962mu}
\begin{eqnarray}
V = \left(\matrix{ V^{}_{e1} & V^{}_{e2} & V^{}_{e3}\cr
V^{}_{\mu 1} & V^{}_{\mu 2} & V^{}_{\mu 3}\cr
V^{}_{\tau 1} & V^{}_{\tau 2} & V^{}_{\tau 3} }\right) \; ,
\end{eqnarray}
which is a $3\times 3$ unitary matrix. It is usually parametrized in terms of three mixing angles $\{\theta^{}_{12}, \theta^{}_{23}, \theta^{}_{13}\}$ and one CP-violating phase $\delta$~\cite{pdg}. If neutrinos are Majorana particles, two additional CP-violating phases $\{\rho, \sigma\}$ corresponding to two relative phases of three neutrino mass eigenstates are needed. The latest global-fit analysis of neutrino oscillation data yields~\cite{Esteban:2016qun}
\begin{eqnarray}
|V^{}_{e1}| \in [0.800, 0.844] \;,~ |V^{}_{e2}| \in [0.515, 0.581] \;, && \nonumber \\
|V^{}_{\mu1}| \in [0.229, 0.516] \;,~ |V^{}_{\tau 1}| \in [0.249, 0.528] \;, && \nonumber \\
|V^{}_{\mu2}| \in [0.438, 0.699] \;,~ |V^{}_{\tau 2}| \in [0.462, 0.715] \;, && \nonumber \\
|V^{}_{\mu3}| \in [0.614, 0.790] \;,~ |V^{}_{\tau 3}| \in [0.595, 0.776] \;, &&
\end{eqnarray}
and $|V^{}_{e3}| \in [0.139, 0.155]$ at the $3\sigma$ level, where the conditions of unitarity have been imposed. Although $\Delta m^2_{21} \approx 7.5 \times 10^{-5}~{\rm eV}^2$ and $|\Delta m^2_{31}| \approx 2.5 \times 10^{-3}~{\rm eV}^2$ have been determined, it remains unknown whether neutrino masses take a normal ordering $m^{}_1 < m^{}_2 < m^{}_3$ (NO, i.e., $\Delta m^2_{31} > 0$) or an inverted ordering $m^{}_3 < m^{}_1 < m^{}_2$ (IO, i.e., $\Delta m^2_{31} < 0$). In addition, the CP-violating phase $\delta$ is not yet measured, though a weak hint for $\delta \approx 270^\circ$ has been observed~\cite{Esteban:2016qun}.

Apart from neutrino oscillations, the $\beta$ decays and neutrinoless double-beta ($0\nu2\beta$) decays could also offer useful information on the PMNS matrix and the absolute neutrino mass. In $\beta$ decays, a finite neutrino mass affects the electron spectrum near its end point, and thus the effective neutrino mass
\begin{eqnarray}
m^{}_\beta \equiv \sqrt{m^2_1 |V^{}_{e1}|^2 + m^2_2 |V^{}_{e2}|^2 + m^2_3 |V^{}_{e3}|^2} \; ,
\end{eqnarray}
can be extracted from the distorted spectrum. Current upper limit $m^{}_\beta < 2.2~{\rm eV}$ at the $95\%$ confidence level (CL) has been obtained from the tritium beta decays in the Mainz and Troitsk experiments~\cite{Kraus:2004zw,Aseev:2011dq}. In the near future, the KATRIN experiment will push such a limit down to $m^{}_\beta < 0.2~{\rm eV}$~\cite{Wolf:2008hf}. If massive neutrinos are Majorana particles, the $0\nu2\beta$ decays $N(Z, A) \to N(Z+2, A) + 2e^-$ could take place for some even-even nuclei, such as ${^{76}}{\rm Ge}$ and ${^{136}}{\rm Xe}$~\cite{Rodejohann:2011mu,Bilenky:2014uka}. So far, there is no clear signal of $0\nu2\beta$ decays, implying a restrictive upper bound on the effective neutrino mass
\begin{eqnarray}
m^{}_{\beta \beta} \equiv \left|V^2_{e1} m^{}_1 + V^2_{e2} m^{}_2 + V^2_{e3} m^{}_3\right| \; .
\end{eqnarray}
After the Phase-II running, the KamLAND-Zen Collaboration has achieved the most stringent bound on the half-life of ${^{136}{\rm Xe}}$ in the $0\nu2\beta$ decay mode $T^{1/2}_{0\nu} > 1.07\times 10^{26}~{\rm yr}$ at $90\%$ CL~\cite{KamLAND-Zen:2016pfg}, corresponding to $m^{}_{\beta \beta} < (0.061 \cdots 0.165)~{\rm eV}$, where the uncertainty arises from the calculation of nuclear matrix elements. For Majorana neutrinos in the IO case, the next-generation experiments at the ton scale will be able to discover $0\nu2\beta$ decays~\cite{Bilenky:2014uka}.

Finally, the observations of cosmic microwave background and large-scale structures constrain the sum of three neutrino masses $\Sigma \equiv m^{}_1 + m^{}_2 + m^{}_3$. In the standard $\Lambda$CDM cosmology, the latest result from Planck Collaboration shows $\Sigma < 0.23~{\rm eV}$ at $95\%$ CL~\cite{Ade:2015xua}.

\section{Selected Topics}
\label{select}

\subsection{Phenomenology of neutrino oscillations}

Besides precision measurements, the primary goals of future oscillation experiments are to pin down neutrino mass ordering and discover leptonic CP violation. To probe neutrino mass ordering, there are in general two practical ways:
\begin{itemize}
\item {\it Oscillations in vacuun} -- As first noticed by Petcov and Piai~\cite{Petcov:2001sy}, it is possible to determine neutrino mass ordering via precise measurement of reactor neutrinos at a medium baseline if $\theta^{}_{13}$ is relatively large. The basic idea is an interference between $\Delta m^2_{21}$- and $\Delta m^2_{31}$-driven oscillations. Experiments implementing this idea include JUNO~\cite{Zhan:2008id,Zhan:2009rs,Li:2013zyd,An:2015jdp} and RENO-50~\cite{Kim:2014rfa}.

\item {\it Matter effects} -- As realized by Wolfenstein~\cite{Wolfenstein:1977ue}, Mikheev and Smirnov~\cite{Mikheev:1986gs}, neutrino oscillations can be significantly modified by the medium, in which neutrinos are propagating. Such an MSW effect can be characterized by $A \equiv 2\sqrt{2} G^{}_{\rm F} N^{}_e E$, where $G^{}_{\rm F}$ is the Fermi constant, $N^{}_e$ the net electron number density, and $E$ the neutrino beam energy. As a consequence, the oscillation probabilities of neutrinos and antineutrinos involve some terms like $1 \mp A/\Delta m^2_{31}$, which are sensitive to the relative sign between $\Delta m^2_{31}$ and $A$. This feature has been taken up in the long-baseline accelerator experiments T2K~\cite{Abe:2011ks}, NO$\nu$A~\cite{Patterson:2012zs} and DUNE~\cite{Acciarri:2015uup}, and the future huge atmospheric neutrino experiments PINGU~\cite{Aartsen:2014oha}, ORCA~\cite{Katz:2014tta}, Hyper-Kamiokande~\cite{Abe:2011ts} and INO~\cite{Ahmed:2015jtv}.
\end{itemize}
All the experiments in the second category are also sensitive to leptonic CP violation. In this regard, two low-energy neutrino super-beam experiments, namely, ESS$\nu$SB~\cite{Baussan:2013zcy,Wildner:2016dst} and MOMENT~\cite{Cao:2014bea,Blennow:2015cmn}, can reach a competitive sensitivity.

Since a number of oscillation experiments are under construction and will be built, it is interesting and timely to study the phenomenology of neutrino oscillations relevant for one or more experiments.  For instance, the terrestrial matter effects on JUNO and RENO-50 have been investigated in Ref.~\cite{Li:2016txk}. The JUNO will use a 20 kiloton liquid-scintillator detector with an energy resolution of $3\%$ for $E = 1~{\rm MeV}$ in order to distinguish between the spectral distortions for NO and IO. Such an experimental setup renders it possible to achieve a precision below $1\%$ for both $\sin^2 2\theta^{}_{12}$ and $\Delta m^2_{21}$~\cite{An:2015jdp}. For reactor antineutrinos, the effective neutrino mass-squared difference $\Delta \tilde{m}^2_{21}$ and mixing angle $\tilde{\theta}^{}_{12}$ in matter are given by~\cite{Li:2016txk}
\begin{eqnarray}
\Delta \tilde{m}^2_{21} &\approx& \Delta m^2_{21} (1 + \hat{A} \cos 2\theta^{}_{12}) \; , \nonumber \\
\sin^2 2\tilde{\theta}^{}_{12} &\approx& \sin^2 2\theta^{}_{12} (1 - 2 \hat{A} \cos 2\theta^{}_{12}) \; ,
\end{eqnarray}
where $\hat{A} \equiv A/\Delta m^2_{21}$ with $A \approx 7.9\times 10^{-7}~{\rm eV}^2~{E/(4~{\rm MeV})}$ and $\Delta m^2_{21} = 7.5\times 10^{-5}~{\rm eV}^2$. It is obvious that the matter corrections to mixing parameters are measured by $\hat{A}$ and can be as large as $1\%$, which should be taken into account for future data analysis. However, it has been found that the JUNO sensitivity to neutrino mass ordering is slightly reduced by $\Delta \chi^2 \approx 0.5$ when the matter effects are considered~\cite{Li:2016txk}.

\subsection{Lepton Flavor Mixing}

An important theoretical issue is how to understand the flavor mixing pattern. For this purpose, two different but correlative methods are usually adopted, i.e., flavor symmetries and flavor textures.

In the former approach, one extends the particle content of the standard model (SM) and imposes a flavor symmetry on the generic lagrangian. Thus, the flavor mixing pattern is constrained by the flavor symmetries and receives corrections from symmetry breaking. See, e.g., Ref.~\cite{King:2013eh}, for a recent review on discrete flavor symmetries. In the latter approach, one conjectures specific flavor textures of lepton mass matrices, which might also be realized via flavor symmetries or other dynamical mechanisms.

Looking at the allowed ranges of the PMNS matrix elements in Eq.~(2), especially the last three lines, one can observe that a $\mu$-$\tau$ exchange symmetry~\cite{Xing:2008fg}
\begin{eqnarray}
|V^{}_{\mu i}| = |V^{}_{\tau i}| \; , ~ {\rm for} ~ i = 1, 2, 3
\end{eqnarray}
seems to hold. In the standard parametrization~\cite{pdg}, such a symmetry implies~ (1) $\theta^{}_{23} = 45^\circ$ and $\theta^{}_{13} = 0$ or (2) $\theta^{}_{23} = 45^\circ$ and $\delta = \pm 90^\circ$. Obviously, the first choice has been excluded by the result of $\theta^{}_{13} \approx 9^\circ$ from Daya Bay, whereas the second one survives. Future oscillation experiments will further test these predictions, and some deviations from $\theta^{}_{23} = 45^\circ$ or $\delta = \pm 90^\circ$ may point to a partial $\mu$-$\tau$ symmetry, namely, either $|V^{}_{\mu 1}| = |V^{}_{\tau 1}|$ or $|V^{}_{\mu 2}| = |V^{}_{\tau 2}|$~\cite{Xing:2014zka}.

In the basis where the charged-lepton mass matrix $M^{}_l = {\rm diag}\{m^{}_e, m^{}_\mu, m^{}_\tau\}$ is diagonal, the $\mu$-$\tau$ exchange symmetry in the PMNS matrix $V$ can be promoted to a symmetry in the Majorana neutrino mass matrix $M^{}_\nu = V \cdot {\rm diag}\{m^{}_1, m^{}_2, m^{}_3\} \cdot V^{\rm T}$, which is invariant under the following transformation
\begin{eqnarray}
\nu^{}_{e{\rm L}} \to \nu^{}_{e{\rm L}} \; ,~~ \nu^{}_{\mu {\rm L}} \to \nu^{}_{\tau {\rm L}} \; ,~~ \nu^{}_{\tau {\rm L}} \to \nu^{}_{\mu {\rm L}} \; ,
\end{eqnarray}
where $\nu^{}_{\alpha {\rm L}}$ denotes the left-handed neutrino field for $\alpha = e, \mu, \tau$. In addition, the invariance of the Majorana neutrino mass term under another transformation
\begin{eqnarray}
\nu^{}_{e{\rm L}} \to \nu^{\rm C}_{e{\rm L}} \; ,~~ \nu^{}_{\mu {\rm L}} \to \nu^{\rm C}_{\tau {\rm L}} \; ,~~ \nu^{}_{\tau {\rm L}} \to \nu^{\rm C}_{\mu {\rm L}} \; ,
\end{eqnarray}
with $\nu^{\rm C}_{\alpha {\rm L}}$ being the charge conjugate of $\nu^{}_{\alpha{\rm L}}$, has been called the $\mu$-$\tau$ reflection symmetry~\cite{Harrison:2002et,Harrison:2004he,Grimus:2003yn} or generalized $\mu$-$\tau$ symmetry~\cite{Zhou:2012zj}.

Since the strong mass hierarchy $m^{}_\mu \ll m^{}_\tau$ indicates a serious breakdown of any $\mu$-$\tau$ flavor symmetry in the charged-lepton sector, the symmetry defined by Eq.~(7) or Eq.~(8) or that even generalized to $\nu^{}_{\alpha {\rm L}} \to X^{}_{\alpha \beta} \nu^{\rm C}_{\beta {\rm L}}$ (with $X$ being a unitary matrix in the flavor space) can be regarded as a residual symmetry only for neutrinos. The full flavor symmetry will be broken spontaneously or explicitly into a $Z^{}_3$ symmetry for charged leptons, and a generalized $\mu$-$\tau$ flavor symmetry for neutrinos. For a recent review on the $\mu$-$\tau$ flavor symmetry, see Ref.~\cite{Xing:2015fdg}.

Without making use of any flavor symmetries, one can consider simple but viable textures of lepton mass matrices, which will be soon verified or ruled out by precision data of neutrino oscillations. In the literature, the Majorana neutrino mass matrix $M^{}_\nu$ with two texture zeros has been extensively discussed ~\cite{Frampton:2002yf,Xing:2002ta,Fritzsch:2011qv,Zhou:2015qua}. Take the texture ${\bf B}^{}_4$ for example,
\begin{eqnarray}
M^{}_\nu = \left(\matrix{a & b & {\bf 0} \cr
b & c & d \cr
{\bf 0} & d & {\bf 0}}\right)\; ,
\end{eqnarray}
one can easily establish the relationship between neutrino mass ratios and the PMNS matrix elements~\cite{Zhou:2015qua}
\begin{eqnarray}
\frac{m^{}_1}{m^{}_3} e^{2{\rm i}\rho} = \frac{V^*_{\mu 1} V^{}_{\tau 3}}{V^*_{\mu 3} V^{}_{\tau 1}} \; , ~~ \frac{m^{}_2}{m^{}_3} e^{2{\rm i}\sigma} = \frac{V^*_{\mu 2} V^{}_{\tau 3}}{V^*_{\mu 3} V^{}_{\tau 2}} \; ,
\end{eqnarray}
where $\rho$ and $\sigma$ are Majorana CP-violating phases. A salient feature of all two-zero textures is that neutrino mass spectrum and CP-violating phases can be expressed in terms of five observables, i.e., three mixing angles and two neutrino mass-squared differences~\cite{Xing:2002ta,Fritzsch:2011qv}. For ${\bf B}^{}_4$, neutrino masses can be fully figured out
\begin{eqnarray}
&& m^{}_3 \approx \sqrt{\Delta m^2_{31}/(1-\cot^4 \theta^{}_{23})} \approx 0.1~{\rm eV}\; , \nonumber \\
&& m^{}_1 \approx m^{}_2 \approx m^{}_3 \cot^2 \theta^{}_{23} \approx 0.087~{\rm eV} \; ,
\end{eqnarray}
where $\theta^{}_{23} \approx 47^\circ$ and $\Delta m^2_{31} = 2.58\times 10^{-3}~{\rm eV}^2$ have been input. In addition, in the limit of $\delta \to \pm 90^\circ$, one can obtain $\theta^{}_{23} \to 45^\circ$ and $m^{}_i \to \infty$, which turns out to be in contradiction with the cosmological bound. Therefore, a nearly-degenerate mass spectrum and almost maximal CP-violating phase $\delta$ will be soon tested in the next-generation neutrino experiments. As shown in Ref.~\cite{Zhou:2015qua}, the flavor structure of $M^{}_\nu$ in Eq.~(9) can be obtained by implementing the non-Abelian flavor symmetry $A^{}_4$ in the minimal type-(I+II) seesaw model~\cite{Gu:2006wj}.

\subsection{Majorana vs. Dirac}

We proceed with another important issue: whether massive neutrinos are their own antiparticles, i.e., Majorana particles~\cite{Majorana:1937vz}. As mentioned before, if massive neutrinos are of Majorana nature, one can in principle observe the $0\nu2\beta$ decays, and the decay rates are governed by the relevant nuclear matrix elements and the effective neutrino mass $m^{}_{\beta \beta}$, which depends on neutrino mixing angles $\{\theta^{}_{12}, \theta^{}_{23}, \theta^{}_{13}\}$, two Majorana CP-violating phases $\{\rho, \sigma\}$ and absolute neutrino masses $\{m^{}_1, m^{}_2, m^{}_3\}$. In the IO case, there will be a lower bound $m^{}_{\beta \beta} > 0.015~{\rm eV}$, no matter how small the lightest neutrino mass is. In the NO case, some subtle cancellation due to two Majorana phases may happen in $m^{}_{\beta \beta}$, leading to an abyss around $m^{}_1 \sim 10^{-3}~{\rm eV}$. In this unlucky region, it will be hopeless to see $0\nu2\beta$ decays even in the far future.

The importance of experimental observation of $0\nu 2\beta$ decays can be made transparent via the Schechter-Valle theorem~\cite{Schechter:1981bd}: if $0\nu2\beta$ decays take place and no intricate cancellations occur, there must exist a Majorana mass term for neutrinos. The quantitative evaluation of the Schechter-Valle theorem has first been carried out in Ref.~\cite{Duerr:2011zd}, and more recently in Ref.~\cite{Liu:2016oph}. The basic strategy is to assume that the $0\nu2\beta$ decays are mediated by heavy particles, which will be integrated out and induce higher-dimensional operators at low energies~\cite{Pas:2000vn}. Such short-range operators should be made of the fermion fields involved in $0\nu2\beta$ decays, namely, light quarks $\{u, d\}$ and electron $e$. The most general Lorentz-invariant operators have been given in Ref.~\cite{Pas:2000vn} and the corresponding $0\nu2\beta$ decay rates have been calculated. For the following operator
\begin{eqnarray}
{\cal O}^{}_{0\nu2\beta} = (2m^{}_p)^{-1} G^2_{\rm F} \epsilon^{\rm xyz}_3 J^\mu_{\rm x} J^{}_{{\rm y} \mu} j^{}_{\rm z} \; ,
\end{eqnarray}
where ${\rm x, y, z}$ denotes the chirality ${\rm L}$ or ${\rm R}$, and the hadronic and leptonic currents are defined as $J^\mu _{{\rm L}/{\rm R}} \equiv \overline{u}\gamma^\mu(1\mp \gamma^{}_5)d$ and $j^{}_{{\rm L}/{\rm R}} \equiv \overline{e}(1\mp \gamma^{}_5)e^{\rm C}$, current lower bound on the half-life $T^{1/2}_{0\nu} > 10^{26}~{\rm yr}$ can be translated into an upper bound $\epsilon^{{\rm R}{\rm R}{\rm L}}_3 < 10^{-8}$. On the other hand, one can draw a four-loop self-energy diagram for light neutrinos by connecting the external fermion lines in $0\nu2\beta$ decays with weak gauge bosons. The Majorana neutrino mass turns out to be~\cite{Duerr:2011zd,Liu:2016oph}
\begin{eqnarray}
(\delta M^{}_\nu)^{}_{ee} \lesssim 10^{-28}~{\rm eV} \; ,
\end{eqnarray}
which is too small to account for neutrino oscillations. Some comments on this result are in order~\cite{Duerr:2011zd,Liu:2016oph}:
\begin{itemize}
\item The Schechter-Valle theorem is qualitatively correct that neutrinos do have a small Majorana mass term if $0\nu2\beta$ decays are observed.

\item Only the short-range contributions to $0\nu 2\beta$ decays have been taken into account in the above evaluation. However, the long-range contributions may even dominate $0\nu 2\beta$ decays.

\item The exchange of light neutrinos might not be the dominant mechanism for $0\nu 2\beta$ decays. This happens when neutrinos are pseudo-Dirac particles.
\end{itemize}
Similarly, one can compute the radiatively-generated neutrino masses from lepton-number-violating (LNV) decays of mesons, assuming the latter are caused by the higher-dimensional operators~\cite{Liu:2016oph}.

If the true value of $m^{}_{\beta \beta}$ falls into the unlucky region, namely, $m^{}_{\beta \beta} \to 0$, how can we probe the Majorana nature of massive neutrinos? The recent experimental proposal PTOLEMY~\cite{Betts:2013uya}, which is designed for detecting cosmic neutrino background via $\nu^{}_e + {^3{\rm H}} \to {^3}{\rm He} + e^-$, gives rise to a new possibility. It has been pointed out in Ref.~\cite{Long:2014zva} that for a $100~{\rm g}$ of surface-deposited tritium source the capture rate will be $4~{\rm yr}^{-1}$ for Dirac neutrinos, while $8~{\rm yr}^{-1}$ twice larger for Majorana neutrinos.

\subsection{Seesaw models}

We come to the origin of neutrino masses. Although it is not yet excluded that neutrinos are Dirac particles, some theoretical difficulties may arise in this case. First of all, the observed sub-eV neutrino masses require extremely small neutrino Yukawa couplings, namely, $y^{}_{\nu_i}/y^{}_t \lesssim 10^{-12}$, where $y^{}_t \sim {\cal O}(1)$ is the top-quark Yukawa coupling. This exaggerates the strong hierarchy problem of fermion masses. Second, right-handed neutrinos $\nu^{}_{\rm R}$'s are singlets under the SM gauge symmetries, so an additional $U(1)$ symmetry has to be introduced to forbid a Majorana mass term of $\nu^{}_{\rm R}$'s.

For Majorana neutrinos, three types of seesaw models can be constructed:
\begin{itemize}
\item {\it Type-I Seesaw} -- Three singlet right-handed neutrinos $\nu^{}_{\rm R}$'s are added into the SM, and a large Majorana mass term for them is allowed. The gauge-invariant lagrangian relevant for neutrino masses can be written as~\cite{type1}
    \begin{eqnarray}
    -{\cal L}^{}_{\rm I} = \overline{\ell^{}_{\rm L}}Y^{}_\nu \tilde{H} \nu^{}_{\rm R} + \frac{1}{2} \overline{\nu^{\rm C}_{\rm R}} M^{}_{\rm R} \nu^{}_{\rm R} + {\rm h.c.} \; ,
    \end{eqnarray}
    where $\tilde{H} \equiv {\rm i}\sigma^{}_2 H^*$ and $\ell^{}_{\rm L}$ stand for Higgs and lepton doublets, respectively. The Majorana mass of light neutrinos is then given by $M^{}_\nu = - v^2 Y^{}_\nu M^{-1}_{\rm R} Y^{\rm T}_\nu$ with $v \equiv \langle H\rangle \approx 174~{\rm GeV}$ being the vacuum expectation value (vev) of Higgs field.
\item {\it Type-II Seesaw} -- One Higgs triplet $\Delta$ is introduced to the SM, and simultaneously coupled to both lepton and Higgs doublets. Thus, the gauge-invariant lagrangian reads~\cite{type2}
    \begin{eqnarray}
    -{\cal L}^{}_{\rm II} = \frac{1}{2} \overline{\ell^{}_{\rm L}} Y^{}_\Delta {\rm i}\sigma^{}_2 \Delta \ell^{\rm C}_{\rm L} + {\rm h.c.} \; .
    \end{eqnarray}
    In the scalar potential $V(H, \Delta)$, the crossing coupling term $\mu^{}_\Delta H^{\rm T} {\rm i}\sigma^{}_2 \Delta H$ is crucial for $\Delta$ to acquire a small vev $v^{}_\Delta \approx \mu^{}_\Delta v^2/M^2_\Delta$, where $M^{}_\Delta$ is the mass of Higgs triplet. Therefore, neutrino masses are given by $M^{}_\nu = Y^{}_\Delta v^{}_\Delta \approx Y^{}_\Delta \mu^{}_\Delta v^2/M^2_\Delta$.

\item {\it Type-III Seesaw} -- Three fermion triplets $\Sigma^{}_{\rm R}$'s are considered to couple with both lepton and Higgs doublets, just like right-handed neutrinos $\nu^{}_{\rm R}$'s. The corresponding lagrangian is given by~\cite{type3}
    \begin{eqnarray}
    -{\cal L}^{}_{\rm III} = \overline{\ell^{}_{\rm L}} Y^{}_\Sigma \Sigma^{}_{\rm R} \tilde{H} + {\rm h.c.} \; .
    \end{eqnarray}
    In this scenario, the electrically neutral component of $\Sigma^{}_{\rm R}$ behaves like the right-handed neutrino $\nu^{}_{\rm R}$. Thus, neutrino masses can be derived from $M^{}_\nu = - v^2 Y^{}_\Sigma M^{-1}_{\Sigma} Y^{\rm T}_\Sigma$, where $M^{}_\Sigma$ is the mass matrix of Higgs triplets.
\end{itemize}
A common feature of all three seesaw models is that the lightness of neutrinos can be ascribed to the heaviness of the new seesaw particles. That is to say, sub-eV neutrino masses ${\cal O}(M^{}_\nu) \sim 0.1~{\rm eV}$ imply the seesaw scale as high as $10^{14}~{\rm GeV}$, given $v \sim 10^2~{\rm GeV}$.

Although a high-scale seesaw model can accommodate both tiny neutrino masses and the cosmological baryon number asymmetry via leptogenesis~\cite{Fukugita:1986hr}, it may suffer from the gauge hierarchy or fine-tuning problem, which stems from the huge radiative corrections induced by heavy particles to Higgs boson mass. By demanding such corrections $\delta m^2_H \lesssim 0.1~{\rm TeV}^2$ in the type-I seesaw model, we get~\cite{Vissani:1997ys,Abada:2007ux,Xing:2009in,Clarke:2015gwa}
\begin{eqnarray}
{\cal O}(M^{}_{\rm R}) \lesssim 10^7~{\rm GeV} \; .
\end{eqnarray}
To avoid this problem, one can either make use of supersymmetry or just lower the seesaw scale down to TeV.

For the low-scale seesaw models, direct searches for the clear LNV signals induced by heavy seesaw particles are accessible at the CERN LHC. The latest searches have been performed by CMS and ATLAS for type-I~\cite{Khachatryan:2015gha,Aad:2015xaa}, type-II~\cite{Chatrchyan:2012ya,ATLAS:2014kca} and type-III~\cite{Aad:2015dha,Aad:2015cxa} seesaws. The null results have placed useful bounds on the couplings and masses of seesaw particles.

\section{Summary}

It is quite promising that neutrino mass ordering and leptonic CP-violating phase can be determined in the next decade, and all the mixing angles will be measured with a high precision. Then, an immediate question is whether we can pin down the true mechanism for neutrino mass generation and lepton flavor mixing.

Unfortunately, the answer is no. Precision measurements from oscillations experiments are helpful, but not enough. A crucially important step is the observation of $0\nu2\beta$ decays, indicating LNV processes exist and neutrinos are Majorana particles. In addition, the direct searches for new-physics signals at the future large colliders will help identify the final theory of neutrinos.




\bibliographystyle{elsarticle-num}



\end{document}